\newcommand{\beq}{\begin{equation}}
\newcommand{\eeq}{\end{equation}}
\newcommand{\beqar}{\begin{eqnarray}}
\newcommand{\eeqar}{\end{eqnarray}}
\newcommand{\beqml}{\begin{mathletters}}
\newcommand{\eeqml}{\end{mathletters}}
\documentstyle[preprint,aps]{revtex}
\begin{document}
\draft
\title{Fluctuation Dissipation Theorem and The Dynamical Renormalization Group}
\author{Achille Giacometti$^{(1)}$, Amos Maritan$^{(2)}$, Flavio
Toigo$^{(3)}$ and Jayanth R. Banavar$^{(4)}$}
\address{$^{(1)}$ Institut f\"{u}r Festk\"{o}rperforschung
des Forschungszentrums J\"{u}lich, \\
Postfach 1913, D-52425, J\"{u}lich, Germany}
\address{$^{(2)}$International School for Advanced Studies \\
and Istituto Nazionale di Fisica Nucleare INFN , \\
via Beirut 2-4, 34014 Trieste, Italy}
\address{$^{(3)}$Dipartimento di Fisica dell'Universit\`a  and \\
 Istituto Nazionale di Fisica della Materia INFM,\\
via Marzolo 8, 35100 Padova, Italy}
\address{$^{(4)}$Department of Physics and Center for Materials Physics,
104 Davey Laboratory,\\
The Pennsylvania State University, University Park, PA 16802}
\date{\today}
\maketitle
\begin{abstract}
The relation between a recently introduced dynamical real space
renormalization group and the fluctuation-dissipation theorem is
discussed. An apparent incompatibility is pointed out
and resolved.
\end{abstract}
\narrowtext
Over the years, Langevin dynamics has been proven to be a successful
theoretical approach to non-equilibrium problems such
as critical dynamics \cite{HOH}, growth processes \cite{Godreche},
and interface dynamics \cite{Droz}.
In a previous paper \cite{GMTB} an exact dynamical real space
renormalization group analysis of Langevin dynamics
derivable from a Gaussian field theory was presented.
The method has the same range of applicability as the
static counterpart, namely, one-dimensional
lattices and a whole set of hierarchical or self-similar structures.
Apart from the methodological interest of the scheme {\em per se},
the results of the method are expected to be of practical interest
as well, in view of the realization of
growth processes on electrochemical electrodes \cite{NP}.
Unlike other real space approximate schemes \cite{NG}, one important
feature of this approach is that it contains the static
analog as a particular case. This naturally raises the
question of the compatibility of the scheme with the standard
fluctuation-dissipation theorem \cite{Gardiner} which relates
the mobility (assumed unity in the rest of the paper)
to the variance of the noise which drives the
system. In this note, we will address this issue within the framework of
an exactly solvable model. This model, although simple, is important since it
is the zeroth order one in any perturbative expansion.

Consider a system described by a Gaussian field  theory in 1-d with
Hamiltonian (or action) $H(\{ \varphi \})$ given by:
\beqar \label{gaussian}
H(\{ \varphi \})&=& \frac{1}{2} \sum_{x,y} \varphi_x A_{x,y} \varphi_y
\eeqar
with the field variables $\varphi_x$ defined on the sites $x$ of
a lattice. As it is well known \cite{GM}, the implementation of an {\it exact}
static real space renormalizaton group analysis is possible
only in the case when the matrix $A_{x,y}$ has a
nearest-neighbour restriction:
\beqar \label{Amatrix}
A_{x,y} &=& a_x \delta_{x,y}-\delta_{|x-y|,1}
\eeqar
The renormalization procedure has three basic steps :
\begin{description}
\item{1)} The set $E$ of lattice points is divided into
$E_s$ (sites which survive the decimation) and $E_d$
(sites which are decimated) so that their union is $E$ and their
intersection is null. The degrees of freedom corresponding
to $E_d$ are then eliminated either by integration over the
corresponding field variables, or by elimination of  the corresponding
equations of motion (in the dynamical case).
\item{2)} The surviving fields are rescaled so that the
new Hamiltonian has the same form as the original one.
\item{3)} Lengths are scaled in such a way that the original
lattice constant is recovered.
\end{description}
As result of this procedure, a mapping between the original and the new
set of parameters is obtained.

Let us now recall the results of the same procedure in dynamics \cite{GMTB}.
The simplest Langevin dynamics can be constructed as:
\beqar \label{langevin}
{\partial \varphi_x (t) \over \partial t}  &=& - {\delta H (\{\varphi\})
\over \delta \varphi_x} + \eta_x(t)
= - \sum_{y} A_{x,y} \varphi_y(t) +\eta_x(t).
\eeqar
where the stochastic noise $\eta_x(t)$ is chosen from a Gaussian distribution
which has a zero average and variance
\beqar \label{variance}
<\eta_x(t_1) \eta_y(t_2)> = 2 D_{x,y} \delta(t_1-t_2).
\eeqar
The renormalization scheme works along the same lines as in the
static case, but with the important difference that the
renormalization of the noise has also to be taken into account.
It was shown in ref \cite{GMTB} that a necessary and sufficient
condition for the implementation of the renormalization procedure
was that the matrix
$D_{x,y}$ has  the form:
\beqar \label{matrix d}
D_0 \delta_{x,y} + D_1 \delta_{|x-y|,1}.
\eeqar
Since both $D_0$ and $D_1$ are different from zero under renormalization,
the minimum parameter space for the fluctuation matrix $D$
is given by $\{D_0,D_1\}$, implying nearest-neighbour correlation of the noise.
We stress once again that both the static and the dynamic schemes
we are dealing with are {\em exact}, i.e. closed in the parameter space.

We now turn to the connection with the fluctuation-dissipation theorem.
The Fokker-Planck equation associated with the Langevin dynamics
(\ref{langevin}) with a noise whose variance is (\ref{variance}) is
\cite{Gardiner}:
\beqar \label{FP}
\frac{\partial}{\partial t} P(\{ \varphi\},t) &=& \sum_x
\frac{\delta}{\delta \varphi_x} [ P(\{ \varphi\},t)
\frac{\delta}{\delta \varphi_x}
H(\{ \varphi \})  + \sum_y D_{x,y}
\frac{\delta}{\delta \varphi_y} P(\{ \varphi\},t)].
\eeqar
It is then easy to show from (\ref{FP}) that, if we denote by
$P_{\mbox{eq}} (\{\varphi\})$ the equilibrium probability distribution
obtained in the $t \rightarrow \infty$ limit, a necessary condition for
(${\cal N}^{-1}$ is a normalization factor)
\beqar \label{equilibrium}
P_{\mbox{eq}} (\{\varphi\}) &=& {\cal N}^{-1} e^{-H(\{\varphi\})}
\eeqar
to be satisfied is that the matrix $D_{x,y}$ has a  {\it diagonal} form,
so that in the case of eq. (\ref{matrix d}) $D_1=0$.
Indeed (\ref{langevin}), (\ref{variance}) and (\ref{matrix d})
with $D_1=0$ lead to the equilibrium distribution given by
(\ref{equilibrium}).
However the dynamical renormalization group {\em violates }, in general, the
fluctuation-dissipation theorem since one has to start
with a matrix with a non-diagonal form and the equilibrium distribution
of (7) is not obtained.

We now argue that this inconsistency is only apparent.
We will show that the $D_1 \neq 0$ case will lead to
an equilibrium distribution
of a more general Gaussian Hamiltonian.  However, we will explicitly
demonstrate that equilibrium correlation functions corresponding to
the two Hamiltonians are characterized by the same spatial decay.

Let us consider a more general Gaussian Hamiltonian:
\beqar \label{gaussian2}
\tilde{H}(\{ \varphi \})&=& \frac{1}{2} \sum_{x,y} \varphi_x
\tilde{A}_{x,y} \varphi_y
\eeqar
where the symmetrix matrix $\tilde{A}_{x,y}$ is {\em not}
necessarily restricted to
nearest-neighbours.  On again using eq. (\ref{FP}), it can be shown readily
that a Langevin equation of the more general form:
\beqar \label{langevin2}
{\partial \varphi_x (t)\over\partial t}  &=& - \sum_y D_{x,y}
{\delta \tilde{H}(\{\varphi\}) \over \delta \varphi_y} + \eta_x(t)
\eeqar
leads to
the equilibrium distribution:
\beqar \label{equilibrium2}
P_{\mbox{eq}} (\{\varphi\}) &=& {\cal N}^{-1} e^{-\tilde{H}(\{\varphi\})}.
\eeqar

Thus if one starts with the model defined by eqs.
(\ref{gaussian}-\ref{variance}) with $D_0=1$
and $D_1=0$, so that the equilibrium distribution is given by eq.
(\ref{equilibrium}), then the Dynamical Renormalization Group (DRG) leaves
the Langevin equation of the form (\ref{langevin}) with $A$ and $D$
given by eqs. (\ref{Amatrix}) and (\ref{variance}) respectively,
with $D_1 \neq 0$ after one RG step.

{}From eqs (8-10), the equilibrium distribution is given by
(\ref{equilibrium2}),
with $\tilde{H}$ given by (\ref{gaussian2}) and
\beqar \label{Atilde}
\tilde{A} &=& {D}^{-1} A            .
\eeqar
$\tilde{A}$ is symmetric since both $D$ and $A$ depend only on
the difference $|x-y|$.
Thus initially $\tilde{A}=A$ and then the DRG flow for $\tilde{H}$
occurs in a wider parameter space than the original one given by
eq. (\ref{Amatrix}) (i.e. $\tilde{A}$ no longer has the form (\ref{Amatrix}) ).
On the contrary a static RG would leave $H$ of the form
(\ref{gaussian}-\ref{Amatrix}).
However, as we show below, these two
Hamiltonians are {\em equivalent} in the sense that the corresponding
correlation functions have the same leading behaviour.
It is noteworthy that a {\em static} Gaussian model with interactions
defined by $\tilde{A}$ appearing in the Hamitonian (\ref{gaussian2})
{\em not} restricted to short range, can be exactly renormalized through the
dynamics, provided that the matrix $D \cdot \tilde{A}$ {\em is} restricted
to nearest-neighbor interactions!

As an explicitly solvable example,
let us consider the simplest  case of a one dimensional lattice with $a_x=a$.
The static recursions are \cite{GM}:
\beqar \label{recursion a}
a' &=& a^2 -2.
\eeqar
The corresponding dynamics leads to the following recursions \cite{GMTB}:
\beqml
\beqar
\alpha^{\prime} (\omega') &=& \alpha^2(\omega) -2 \label{rec:1} \\
D_0^{\prime} &=& \frac{3}{4} D_0 + D_1 + o(\omega) \label{rec:2} \\
D_1^{\prime} &=& \frac{1}{8} D_0 + \frac{1}{2} D_1 +o(\omega) \label{rec:3}
\eeqar
\eeqml
where $\alpha(\omega) \equiv a -i \omega$ and $\omega$ is
the Fourier variable conjugate to time.

 From equation (\ref{rec:1}) it is apparent that the static case
is recovered in the limit $\omega \rightarrow 0$
(i.e. $t \rightarrow \infty$), that is the statics is included
in the dynamics as we mentioned already.

Let us now discuss the two-point correlation function given by:
\beqar \label{2pt}
G_{x,y} &=& < \varphi_x \; \varphi_y> = \frac{1}{Z} \int {\cal D} \varphi
 \; e^{-H(\{\varphi\})} \; \varphi_x \varphi_y = (A^{-1})_{x,y}
\eeqar
where $Z$ is the partition function and ${\cal D} \varphi \equiv
\prod d \varphi_x$. A similar expression holds for $\tilde{G}_{x,y}$
corresponding to the Hamiltonian $\tilde{H}(\{\varphi\})$.

Due to the particular form of $A$ and $D$ and using eq.(\ref{Atilde}),
one can easily see that the $G_{x,y}$ and $\tilde{G}_{x,y}$ are
given by (with the lattice constant set to unity):
\beqar  \label{g1}
G_{x,y} &=& \int_{-\pi}^{+\pi} \frac{dq}{2\pi} \; e^{iq(x-y)}
\frac{1}{a-2 \cos q}
\eeqar
and
\beqar  \label{g2}
\tilde{G}_{x,y} &=& \int_{-\pi}^{+\pi} \frac{dq}{2\pi} \; e^{iq(x-y)}
\frac{D_0+2D_1 \cos q}{a-2 \cos q}  .
\eeqar
The integrals can be carried out exactly and in the $a >2$ case, where the
correlation functions are real, they yield the same coarse-grained
behaviour i.e:
\beqar \label{g3}
G_{x,y}&=& \frac{1}{\sqrt{a^2-4}} e^{-\lambda |x-y|}
\eeqar
(when $|x-y| > 1$) and
\beqar \label{g4}
\tilde{G}_{x,y} &=&  \frac{D_0+D_1 a}{\sqrt{a^2-4}} e^{-\lambda |x-y|}
\eeqar
where we have defined:
\beqar \label{lambda}
\lambda &=&|\ln[ \frac{a}{2}(1-\sqrt{1-\frac{4}{a^2}})]|
\eeqar
This shows that the two Hamiltonians $H$ and $\tilde{H}$ are indeed
equivalent, in the sense that the corresponding correlation functions
decay similarly.

Conversely, one may show that $\tilde{A}$ may be written
in the form (\ref{Atilde})
with $A$ and $D$ given by (\ref{Amatrix}) and (\ref{matrix d})
respectively, provided that, at leading order, it is of the form:
\beqar \label{final}
\tilde{A}_{x,y} &=&  \tilde{a} \delta_{x,y}+\tilde{b} e^{-\mu|x-y|}.
\eeqar

In summary we have discussed the issue of the relationship between
the dynamical renormalization group and the fluctuation-dissipation
theorem, which was prompted by our previous analysis.
We have showed that although at first glance there is a violation
of the fluctuation-dissipation theorem, a more careful analysis
restores its validity.

AG acknowledges support from the HCM program under contract ERB4001GT932058.
This work was supported by NATO, ONR (USA), INFN (Italy), EPSRC (UK),
The Fulbright Foundation and The Donors of the Petroleum Research Fund
administered by The American Chemical Society.


\end{document}